\documentclass{ws-procs9x6}

\begin{document}
\hyphenation{ener-gy ave-ra-ge Ave-ra-ge ge-ne-ra-ted re-la-ti-vi-stic dif-fe-ren-ce in-sensi-ti-ve ki-ne-ma-ti-cal Sy-ste-ma-tic sy-ste-ma-tic ave-ra-ging con-ver-ters con-ver-ter la-bo-ra-to-ry se-con-da-ry
se-con-da-ries si-mu-la-tion si-mu-la-tions do-mi-na-te do-mi-na-tes ta-bu-la-tions
Ta-bu-la-tions Di-stri-bu-tions
di-stri-bu-tions di-stri-bu-tion Di-stri-bu-tion Di-spla-ce-ment di-spla-ce-ment Di-spla-ce-ments Bethe power powers ma-te-rial ma-te-rials
di-spla-ce-ments Li-near li-near Ca-sca-de Ca-sca-des ca-sca-de ca-sca-des
ta-bu-la-tion Ta-bu-la-tion re-pe-ti-ti-ve E-lec-tron E-lec-trons
e-lec-tron e-lec-trons De-tec-tion Pro-duc-tion pro-duc-tion
Re-so-lu-tions Re-so-lu-tion re-so-lu-tions re-so-lu-tion
Ope-ra-tion mi-ni-mum Ener-gy Ener-gies fer-ro-mag-net
fer-ro-mag-nets meta-sta-ble meta-sta-bi-lity con-fi-gu-ra-tion
con-fi-gu-ra-tions expo-nen-tially mo-bi-li-ty- mo-bi-li-ties
tem-pe-ra-tu-re tem-pe-ra-tu-res con-cen-tra-tion con-cen-tra-tions
elec-tro-nic elec-tro-nics STMelec-tro-nics sec-tion Sec-tion
Chap-ter chap-ter theo-ry ap-pro-xi-mation ra-dia-tion Ra-dia-tion
ca-pa-ci-tan-ce approaches tran-sport dispersion Ca-lo-ri-me-try
ca-lo-ri-me-try En-vi-ron-ment En-vi-ron-ments en-vi-ron-ment
en-vi-ron-ments Fur-ther-mo-re do-mi-nant ioni-zing pa-ra-me-ter pa-ra-me-ters
O-sa-ka ge-ne-ral exam-ple Exam-ple ca-vi-ty Ca-vi-ty He-lio-sphe-re
he-lio-sphe-re dis-tan-ce Inter-pla-ne-ta-ry inter-pla-ne-ta-ry
ge-ne-ra-li-zed sol-ving pho-to-sphe-re sym-me-tric du-ring
he-lio-gra-phic strea-ming me-cha-nism me-cha-nisms expe-ri-mental
Expe-ri-mental im-me-dia-tely ro-ta-ting na-tu-rally
ir-re-gu-la-ri-ties o-ri-gi-nal con-si-de-red e-li-mi-na-ting
ne-gli-gi-ble stu-died dif-fe-ren-tial mo-du-la-tion ex-pe-ri-ments
ex-pe-ri-ment Ex-pe-ri-ment Phy-si-cal phy-si-cal in-ve-sti-ga-ted
Ano-de Ano-des ano-de ano-des re-fe-ren-ce re-fe-ren-ces
ap-pro-xi-ma-ted ap-pro-xi-ma-te in-co-ming bio-lo-gi-cal
atte-nua-tion other others eva-lua-ted nu-cleon nu-cleons reac-tion
pseu-do-ra-pi-di-ty pseu-do-ra-pi-di-ties esti-ma-ted va-lue va-lues
ac-ti-vi-ty ac-ti-vi-ties bet-ween Bet-ween dis-cre-pan-cy
dis-cre-pan-cies cha-rac-te-ri-stic
cha-rac-te-ri-stics sphe-ri-cally anti-sym-metric ener-gy ener-gies
ri-gi-di-ty ri-gi-di-ties leaving pre-do-mi-nantly dif-fe-rent
po-pu-la-ting acce-le-ra-ted respec-ti-ve-ly sur-roun-ding
sa-tu-ra-tion vol-tage vol-tages da-ma-ge da-ma-ges be-ha-vior
equi-va-lent si-li-con exhi-bit exhi-bits con-duc-ti-vi-ty
con-duc-ti-vi-ties dy-no-de dy-no-des created Fi-gu-re Fi-gu-res
tran-si-stor tran-si-stors Tran-si-stor Tran-si-stors ioni-za-tion
Ioni-za-tion ini-tia-ted sup-pres-sing in-clu-ding maxi-mum mi-ni-mum
vo-lu-me vo-lu-mes tu-ning ple-xi-glas using de-pen-ding re-si-dual har-de-ning li-quid
know-ledge usage me-di-cal par-ti-cu-lar scat-te-ring ca-me-ra se-cond hea-vier hea-vy trans-axial
con-si-de-ration created Hy-po-the-sis hy-po-the-sis usually inte-ra-ction Inte-ra-ction
inte-ra-ctions Inte-ra-ctions pro-ba-bi-li-ty pro-ba-bi-li-ties
fol-low-ing cor-re-spon-ding e-la-stic readers reader pe-riod pe-riods geo-mag-ne-tic sa-ti-sfac-tory ori-gi-nal-ly}

\clearpage
\setcounter{page}{9}

\begin{center}
Updated version including supplementary material of the article published (http://www.worldscientific.com/doi/pdf/10.1142/9789814329033\_0002 www.worldscientific.com/doi/suppl/10.1142/8014/suppl\_file/8014\_errata.pdf)\\by World Scientific (Singapore) in the Proceedings of the \\12th ICATPP Conference \\ Villa  Olmo (Como, Italy), 7--8 October, 2010.
\end{center}
\vspace{-1.5cm}

\title{Nuclear and Non-Ionizing Energy-Loss for Coulomb Scattered Particles from Low Energy up to Relativistic Regime in Space Radiation Environment}

\author{M.J. Boschini$^{1,2}$, C. Consolandi$^{*,1}$, M. Gervasi$^{1,3}$, S. Giani$^{4}$,\\ D. Grandi$^{1}$, V. Ivanchenko$^{4}$,
S. Pensotti$^{3}$, P.G. Rancoita$^{**,1}$, M. Tacconi$^{1}$}

\address{$^1$\textit{INFN-Milano Bicocca, P.zza Scienza,3 Milano, Italy}\\
$^2$\textit{CILEA Via R. Sanzio, 4 Segrate, MI-Italy}\\
$^3$\textit{Milano Bicocca University, Piazza della Scienza, 3 Milano, Italy} \\
$^4$\textit{CERN, Geneva, 23, CH-1211, Switzerland}\\
$^*$E-mail: cristina.consolandi@mib.infn.it\\
$^{**}$E-mail: piergiorgio.rancoita@mib.infn.it}

\begin{abstract}
In the space environment, instruments onboard of spacecrafts can be affected by displacement damage due to radiation.~The differential scattering cross section for screened nucleus--nucleus interactions - i.e., including the effects due to screened Coulomb nuclear fields -, nuclear stopping powers and non-ionization energy losses are treated from about 50\,keV/nucleon up to relativistic energies.
%
\end{abstract}

\bodymatter

\section{Introduction}
In the space environment near Earth, low energy particles are, for instance, found trapped within the radiation belts and are partially coming from the Sun.~On the other hand, the energies of Galactic Cosmic Rays (GCR) extend up to relativistic range.~Protons are the most abundant, but alpha particles and heavier nuclei are also present (e.g.,~see Sections~4.1.2--4.1.2.5 of Ref.~[1]).~Abundances and energy spectra of GCRs depend on the position inside the solar cavity and are affected by the solar activity.~Above (30--50)\,MeV/nucleon, the dominant radiation
consists of GCRs.~At lower energies, from 1\,MeV/nucleon up to about 30\,MeV/nucleon, one also finds the so-called Anomalous Cosmic Rays (ACRs).~GCRs can reach Earth's magnetosphere and interact with upper layers of the atmosphere.~These interactions produce secondary particles, like for example protons with (10--100)\,MeV energies, which may - in turn - become trapped within the radiation belts.~In addition, during transient phenomena like solar flares and coronal mass ejections, Solar Energetic Particles (SEP) are produced in the energy range from few keV's to GeV's.
\par
All these energetic particles can inflict permanent damages to onboard electronic devices employed in space missions.~While passing through matter, they can lose energy by Coulomb interactions with electrons (\textit{electronic energy-loss}) and nuclei (\textit{nuclear energy-loss}) of the material.~In particular, the nuclear energy-loss - due to screened Coulomb scattering on nuclei of the medium - is relevant for the creation of permanent defects inside the lattice of the material; thus, for instance, it is mostly responsible for the displacement damage which is a cause of degradation of silicon devices.~
\par
The developed model - presented in this article - for screened Coulomb elastic scattering up to relativistic energies is included into Geant4 distribution [2] and is available
with Geant4 version 9.4 (December 2010).
\section{Nucleus--Nucleus Interactions and Screened Coulomb Potentials}
\label{Nucleus_Nucleus_Potentials}
At small distances from the nucleus, the potential energy is a Coulomb potential, while - at distances larger than the Bohr radius - the nuclear field is screened by the fields of atomic electrons.~The interaction between two nuclei is usually described in terms of an interatomic Coulomb potential (e.g., see Section~2.1.4.1 of Ref.~[1] and Section~4.1 of Ref.~[3]), which is a function of the radial distance $r$ between the two nuclei
\begin{equation}\label{eq:VcoulombScreen}
 V(r)=\frac{zZ e^2}{r}\, \,\Psi_{\rm I}(r_{\rm r}),
\end{equation}
where $ez$ (projectile) and $eZ$ (target) are the charges of the
bare nuclei and $\Psi_{\rm I}$ is the \index{Interatomic!screening
function}\textit{interatomic screening function}.~This latter
function depends on the \index{Reduced!radius|see{Radius,
reduced}}\index{Radius!reduced}\textit{reduced radius} $r_{\rm r}$
given by
\begin{equation}\label{reduced_radius}
    r_{\rm r}=\frac{r}{\rm {a_{I}}},
\end{equation}
where $\rm {a_{I}}$ is the so-called \index{Screening!length}\textit{screening
length} (also termed \index{Screening!radius}\textit{screening
radius}).~In the framework of the Thomas--Fermi model of the atom (e.g.,~see Chapters~1 and~2 of~Ref.~[4]) -~thus, following the approach of ICRU~Report~49~(1993) -, a commonly used screening length for $z=1$ incoming particles is that from Thomas--Fermi~(e.g.,~see Refs.~[5,~6])
\begin{equation}\label{T_F_sc_rad}
     {\rm {a_{TF}}}=\frac{C_{\rm TF} \,{\rm {a_{0}}} }{Z^{1/3}},
\end{equation}
and - for incoming particles with $z \geq 2$ - that introduced by~Ziegler, Biersack and Littmark~(1985) (and termed \textit{universal screening length}\footnote[1]{Another screening length commonly used is that from Lindhard and Sharff~(1961) (e.g.,~see Ref.~[8] ; see also Ref.~[9] and references therein):
\[
 {\rm {a_{L}}}=\frac{C_{\rm TF} \,{\rm {a_{0}}} }{\left(z^{2/3}+Z^{2/3}\right)^{1/2}}.
\]})
\begin{equation}\label{Universal_sc_rad}
  {\rm {a_{U}}} = \frac{C_{\rm TF} \,{\rm {a_{0}}}}{z^{0.23}+Z^{0.23}},
\end{equation}
where
\[
{\rm {a_{0}}} =\frac{\hbar^2}{m e^2}
\]
is the \index{Bohr!radius}Bohr radius,
$m$ is the electron rest mass and
\[
C_{\rm TF} = \frac{1}{2}\left(\frac{3 \, \pi}{4} \right)^{2/3}  \simeq 0.88534
\]
is a constant introduced in the
\index{Thomas--Fermi!model}Thomas--Fermi model.~
\par
The simple scattering model due to Wentzel [10] - with a single exponential screening-function $\Psi_{\rm I}(r_{\rm r})$ \{e.g.,~see Ref.~[10] and Equation~(21) in Ref.~[11]\} - was repeatedly employed in treating single and multiple Coulomb-scattering with screened potentials (e.g.,~see Ref.~[11] - and references therein - for a survey of such a topic and also Refs.~[12--15]).~The resulting elastic differential cross section differs from the Rutherford differential cross section by an additional term - the so-called \textit{screening parameter} - which prevents the divergence of the cross section when the angle $\theta$ of scattered particles approaches $0^\circ$.~The screening parameter $A_{\rm s,M}$ [e.g.,~see Equation (21) of Bethe~(1953)] - as derived by Moli\`{e}re~(1947,~1948) for the single Coulomb scattering using a Thomas--Fermi potential - is expressed\footnote[2]{It has to be remarked that the screening radius originally used in~Refs.~[12,~13] was that from Eq.~(\ref{T_F_sc_rad}).} as
\begin{equation}\label{eq:As}
A_{\rm s,M}=\left(\frac{\hbar}{2\,p\ a_{\rm I}}\right)^2\left[1.13+3.76 \times \left(\frac{\alpha zZ}{\beta}\right)^2\right]
\end{equation}
where $a_{\rm I}$ is the screening length - from Eqs.~(\ref{T_F_sc_rad},~\ref{Universal_sc_rad}) for particles with $z =1$ and $z \geq 2$, respectively; $\alpha$ is the fine-structure constant; $p$ ($\beta c$) is the momentum (velocity) of the incoming particle undergoing the scattering onto a target supposed to be initially at rest; $c$ and $\hbar$ are the speed of light and the reduced Planck constant, respectively.~When the (relativistic) mass - with corresponding rest mass $m$ - of the incoming particle is much lower than the rest mass ($M$) of the target nucleus, the differential cross section - obtained from the Wentzel--Moli\`{e}re treatment of the single scattering - is:
\begin{eqnarray}
\label{eq:diff_cross_W_M_model}    \frac{d\sigma^{\rm WM}(\theta)}{d\Omega} & = & \left(\frac{zZe^2}{p \, \beta c }\right)^2\frac{1}{(2A_{\rm s,M}+1-\cos{\theta})^2}\\
\label{eq:diff_cross_W_M_model_sin} &=& \left(\frac{zZe^2}{2\,p \, \beta c }\right)^2\frac{1}{\left[A_{\rm s,M}+\sin^2({\theta}/2)\right]^2}
\end{eqnarray}
(e.g.,~see Section~2.3 in Ref.~[11] and references therein).~Equation~(\ref{eq:diff_cross_W_M_model_sin}) differs from Rutherford's formula - as already mentioned - for the additional term $A_{\rm s,M}$ to $\sin^2({\theta}/2)$.~The corresponding total cross section \{e.g., see Equation~(25) in~Ref.~[11]\} per nucleus is
\begin{equation} \label{eq:cross_W_M_model_t}
\sigma^{\rm WM} =\left(\frac{zZe^2}{p \, \beta c }\right)^2\frac{\pi }{A_{\rm s,M} (1+A_{\rm s,M})}.
\end{equation}
Thus, for $\beta \simeq 1$ (i.e.,~at very large $p$) and with $A_{\rm s,M}\ll 1$, from Eqs.~(\ref{eq:As},~\ref{eq:cross_W_M_model_t}) one finds that the cross section approaches a constant:
\begin{equation} \label{eq:cross_W_M_model_t_asy}
\sigma^{\rm WM}_{\rm c}  \simeq \left(\frac{2\, zZe^2 a_{\rm I}}{\hbar c }\right)^2\frac{  \pi }{1.13+3.76 \times \left(\alpha zZ\right)^2 }.
\end{equation}
\par
In case of a scattering under the action of a central potential (for instance that due to a screened Coulomb field), when the rest mass of the target particle is no longer much larger than the relativistic mass of the incoming particle, the expression of the differential cross section must properly be re-written - in the center of mass system - in terms of an ``effective particle'' with momentum ($p_{\rm r}'$) equal to that of the incoming particle ($p'_{in}$) and rest mass equal to the relativistic reduced mass
\[
\mu_{\rm rel} =   \frac{m M }{M_{1,2}},
\]
where $M_{1,2}$ is the invariant mass; $m$ and $M$ are the rest masses of the incoming and target particles, respectively (e.g., see~Refs.~[15--17] and references therein).~The ``effective particle'' velocity is given by:
\[
\beta_{\rm r} c = c \sqrt{\left[1+ \left(\frac{\mu_{\rm rel} c}{p'_{in} }  \right)^2 \right]^{-1}}.
\]
Thus, the differential cross section\footnote[6]{By inspection of Eqs.~(\ref{eq:As},~\ref{eq:diff_cross_W_M_model_sin},~\ref{eq:diff_cross_W_M_model_r},~\ref{eq:As_r}), one finds that for $\beta_{\rm r} \approxeq 1$ the cross section is given by Eq.~(\ref{eq:cross_W_M_model_t_asy}).} per unit solid angle of the incoming particle results to be given by
\begin{equation}\label{eq:diff_cross_W_M_model_r}
   \frac{d\sigma^{\rm WM}(\theta')}{d\Omega'}  =  \left(\frac{zZe^2}{2\,p'_{in} \, \beta_{\rm r} c }\right)^2\frac{1}{\left[A_{\rm s}+\sin^2({\theta'}/2)\right]^2},
\end{equation}
with
\begin{equation}\label{eq:As_r}
A_{\rm s}=\left(\frac{\hbar}{2\,p'_{in}\ a_{\rm I}}\right)^2\left[1.13+3.76 \times \left(\frac{\alpha zZ}{\beta_{\rm r}}\right)^2\right]
\end{equation}
and $\theta'$ the scattering angle in the center of mass system.~
\par
The energy\footnote[3]{One can show -~e.g.,~see Section~1.5 of Ref.~[1] - that the four momentum transfer is given by
\[
t= - 2M T.
\]
Since $t$ is invariant, then the kinetic energy transferred is also invariant.~Furthermore, since $T=T_{max}\sin^2{(\theta'/2)}$, then one finds that
\[
d [-T] =  T_{max}\,d [-\sin^2{(\theta'/2)}]= \frac{T_{max}}{2}\, \sin{(\theta')} \,d\theta'
\]
(e.g.,~see Section~1.5 of Ref.~[1]).}
$T$ transferred to the recoil target is related to the scattering angle as  $T=T_{max}\sin^2{(\theta'/2)}$ - where $T_{max}$ is the maximum energy which can be transferred in the scattering (e.g.,~see Section~1.5 of Ref.~[1]) -, thus, assuming an isotropic azimuthal distribution one can re-write Eq.~(\ref{eq:diff_cross_W_M_model_r}) in terms of the kinetic energy transferred from the projectile -~,~i.e., [$-T$], where the negative sign indicates that energy is lost by the projectile - to the recoil target as
\begin{eqnarray}
     \nonumber{d\sigma^{\rm WM}}  &=& \left(\frac{zZe^2}{2\,p'_{in} \, \beta_{\rm r} c }\right)^2\frac{1}{\left[A_{\rm s}+\sin^2({\theta'}/2)\right]^2} \sin( \theta')\, d \theta' \int_{0}^{2\pi} d\phi \\
     \label{eq:diff_cross_W_M_model_T_negative} &=&  \pi \left(\frac{zZe^2}{p'_{in} \, \beta_{\rm r} c }\right)^2\frac{T_{max}}{\left[T_{max}\, A_{\rm s}+T\right]^2} \,d [-T].
  \end{eqnarray}
Finally, from Eq.~(\ref{eq:diff_cross_W_M_model_T_negative}), the differential cross section with respect to the kinetic recoil energy ($T$) of the target is given by:
\begin{equation}\label{eq:diff_cross_W_M_model_T}
    \frac{d\sigma^{\rm WM}(T)}{dT} =  \pi \left(\frac{zZe^2}{p'_{in} \, \beta_{\rm r} c }\right)^2\frac{T_{max}}{\left[T_{max}\, A_{\rm s}+T\right]^2}.
\end{equation}
\par
Furthermore, since
\begin{eqnarray}
 \nonumber  \beta_{\rm r} c&=& \frac{p \,c^2 }{E}\\
 \label{pin_p_inc}  p'_{in} &=& \frac{ p M}{M_{1,2}} \\
 \nonumber T_{max}  &=&  \frac{2 p^2 M}{M_{1,2}^2}
\end{eqnarray}
with $p$ and $E$ the momentum and total energy of the incoming particle in the laboratory, then one finds
\[
\frac{T_{max}}{\left(p'_{in} \, \beta_{\rm r} c \right)^2} =\frac{ 2 E^2}{p^2 M c^4}.
\]
Therefore, Eq.~(\ref{eq:diff_cross_W_M_model_T}) can be re-written as
\begin{equation}\label{eq:diff_cross_W_M_model_T_1}
 \frac{d\sigma^{\rm WM}(T)}{dT}  = 2\,  \pi \left({zZe^2}\right)^2 \frac{  E^2}{p^2 M c^4}\, \frac{1}{\left[T_{max}\, A_{\rm s}+T\right]^2}.
\end{equation}
Equation~(\ref{eq:diff_cross_W_M_model_T_1}) expresses - as already mentioned - the differential cross section as a function of the (kinetic) energy $T$ achieved by the recoil target.
\section{Nuclear Stopping Power}
%
\begin{figure}[b]
\begin{center}
\vspace{-1.3cm}
\mbox{\hspace{-0.5cm}\psfig{file=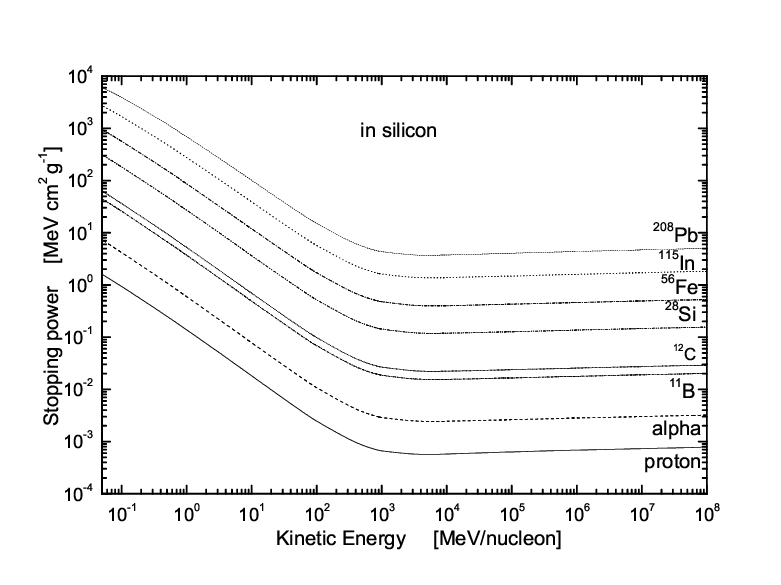,width=4in}}
\end{center}
\vspace{-0.8cm}
\caption{Nuclear stopping power - in MeV\,cm$^2$\,g$^{-1}$ - calculated using Eq.~(\ref{de/dx_nuclear_T}) in silicon is shown as a function of the kinetic energy per nucleon - from 50\,keV/nucleon  up 100\,TeV/nucleon - for protons, $\alpha$-particle and $^{11}$B-, $^{12}$C-, $^{28}$Si-, $^{56}$Fe-, $^{115}$In-, $^{208}$Pb-nuclei.}
\label{fig:dEdx}
\end{figure}
\begin{figure}[t]
\begin{center}
\vspace{-1.3cm}
\mbox{\hspace{-0.5cm} \psfig{file=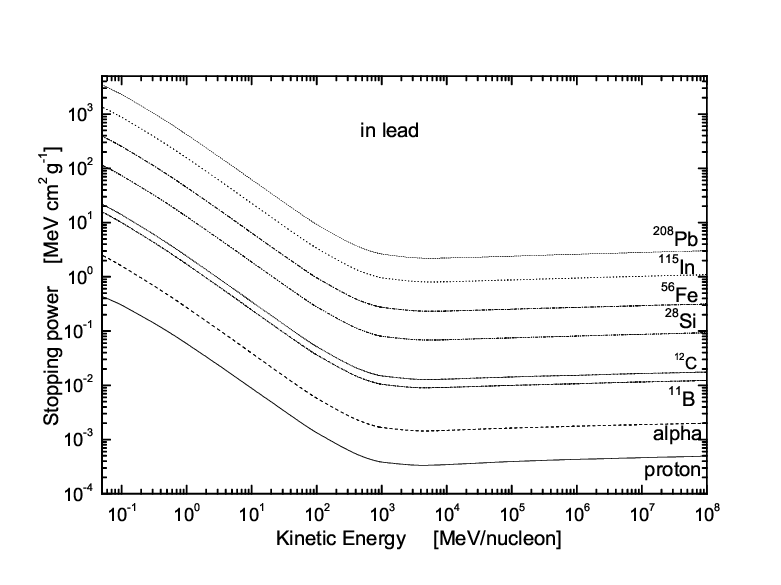,width=4in}}
\end{center}
\vspace{-0.8cm}
\caption{Nuclear stopping power - in MeV\,cm$^2$\,g$^{-1}$ - calculated using Eq.~(\ref{de/dx_nuclear_T}) in lead is shown as a function of the kinetic energy per nucleon - from 50\,keV/nucleon  up 100\,TeV/nucleon - for protons, $\alpha$-particles and $^{11}$B-, $^{12}$C-, $^{28}$Si-, $^{56}$Fe-, $^{115}$In-, $^{208}$Pb-nuclei.}
\label{fig:dEdx_Pb}
\end{figure}
Using Eq.~(\ref{eq:diff_cross_W_M_model_T_1})
the nuclear stopping power - in MeV\,cm$^{-1}$ - is obtained as
\begin{eqnarray}
\label{de/dx_nuclear_WM} - \left(\frac{dE}{dx}\right)_{\rm nucl}  &=& {n_A}\int_{0}^{T_{max}} \frac{d\sigma^{\rm WM}(T)}{dT}\, T \,d T  \\
\nonumber & =&  2\, {n_A} \pi \left({zZe^2}\right)^2 \frac{  E^2}{p^2 M c^4}\!\int_{0}^{T_{max}}\!\frac{T }{\left[A_{\rm s} T_{max}+T\right]^2} \, dT\\
\label{de/dx_nuclear_T} &=&2\, {n_A} \pi \left({zZe^2}\right)^2 \frac{  E^2}{p^2 M c^4}\!\left[\frac{A_{\rm s}}{A_{\rm s}+1}-1+\ln\left(\!\frac{A_{\rm s}+1}{A_{\rm s}}\!\right)\right]
\end{eqnarray}
with $n_A$ the number of nuclei (atoms) per unit of volume and, finally, the negative sign indicates that the energy is lost by the incoming particle (thus, achieved by recoil targets).~For energies higher than a few tens of keVs, because $A_{\rm s} \ll 1$, Eq.~(\ref{de/dx_nuclear_T}) can be re-written as
\begin{equation}\label{de/dx_nuclear_T_ap}
   - \left(\frac{dE}{dx}\right)_{\rm nucl} = 2 \,\pi {n_A}  \left(zZe^2\right)^2 \frac{  E^2}{p^2 M c^4} \left[\ln\left(\!\frac{1}{A_{\rm s}}\!\right)-1\right].
\end{equation}
It has to be noted that, as the incoming momentum increases to a value for which $p \simeq E$, the set of terms - in front of those included in brackets~- decreases and approaches a constant; while the term $\ln\left({1}/{A_{\rm s}}\right)$ increases as $\ln (p)$ for $E \gg m c^2, M c^2$ [e.g.,~see Eqs.~(\ref{eq:As_r},~\ref{pin_p_inc})].~Thus, a slight increase of the nuclear stopping power with energy is expected because of the decrease of the screening parameter with energy.
\begin{figure}[t]
\begin{center}
\vspace{-0.8cm}
\psfig{file=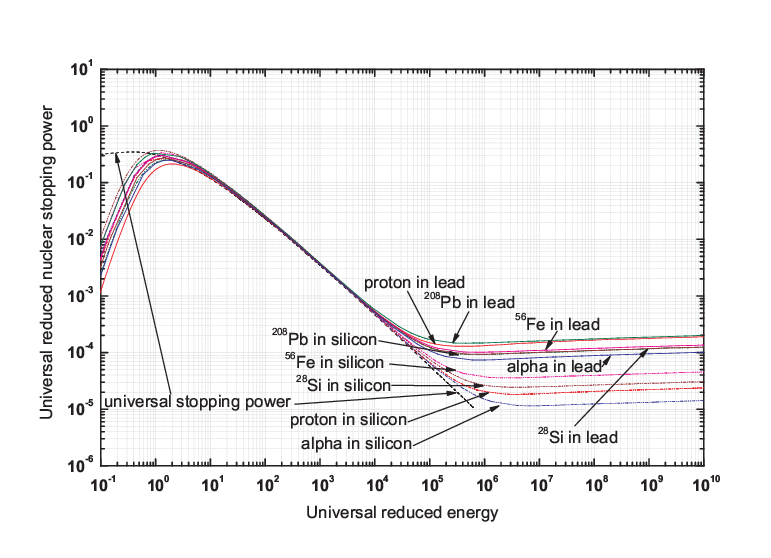,width=4.2in}
\end{center}
\vspace{-0.8cm}
\caption{Universal stopping power (dashed line) as a function of the universal reduced energy [Eq.~(\ref{universal_red_energy})].~The other curves correspond to the dimensionless nuclear stopping power obtained from Eq.~(\ref{de/dx_nuclear_T}) - for protons, $\alpha$-particles, $^{28}$Si-, $^{56}$Fe-, $^{208}$Pb-nuclei in silicon (Fig.~\ref{fig:dEdx}) and lead (Fig.~\ref{fig:dEdx_Pb}) absorbers - and divided by the parameter $\mathbb{K}$ [Eq.~(\ref{practical_NSP_k_PARAMETER})]. }
\label{fig:ZBL}
\end{figure}
\begin{figure}[t]
\begin{center}
\vspace{-0.8cm}
\psfig{file=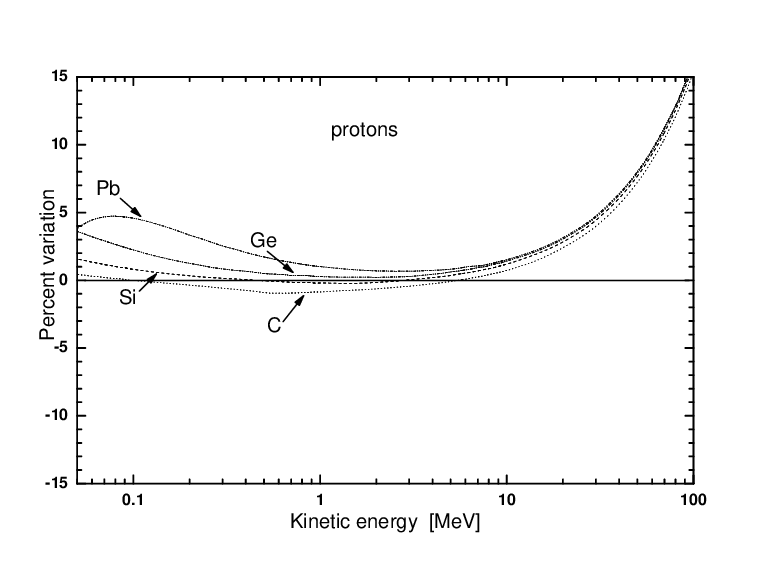,width=2.7in}
\hfil
\vskip -1.0cm
\psfig{file=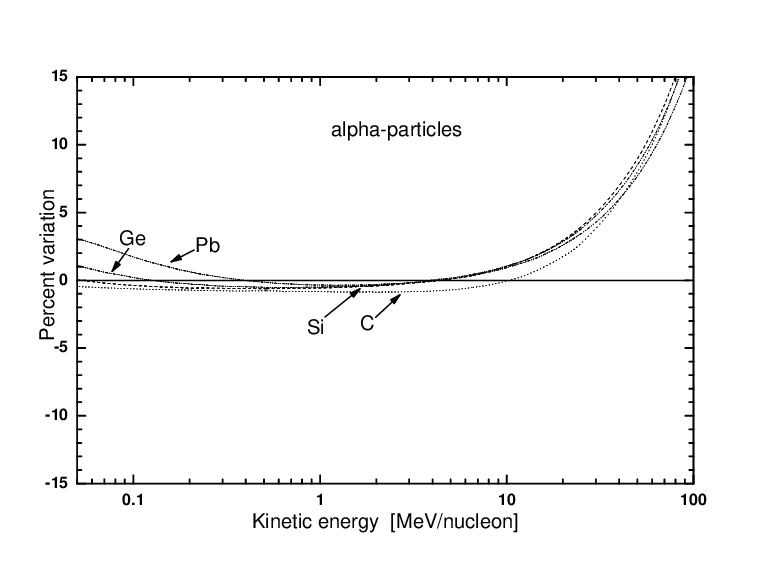,width=2.7in}
\end{center}
\vspace{-0.8cm}
\caption{Variation (in percentage) of nuclear stopping powers - calculated with Eq.~(\ref{de/dx_nuclear_T}) for energies from 50\,keV/nucleon up to 100\,MeV/nucleon - with respect to ICRU tabulated values [3] as a function of the kinetic energy in MeV/nucleon, for protons (top) and $\alpha$-particles (bottom) traversing amorphous carbon, silicon, germanium and lead media.}
\label{fig:Var}
\end{figure}
\begin{figure}[b]
\begin{center}
\vspace{-0.8cm}
\psfig{file=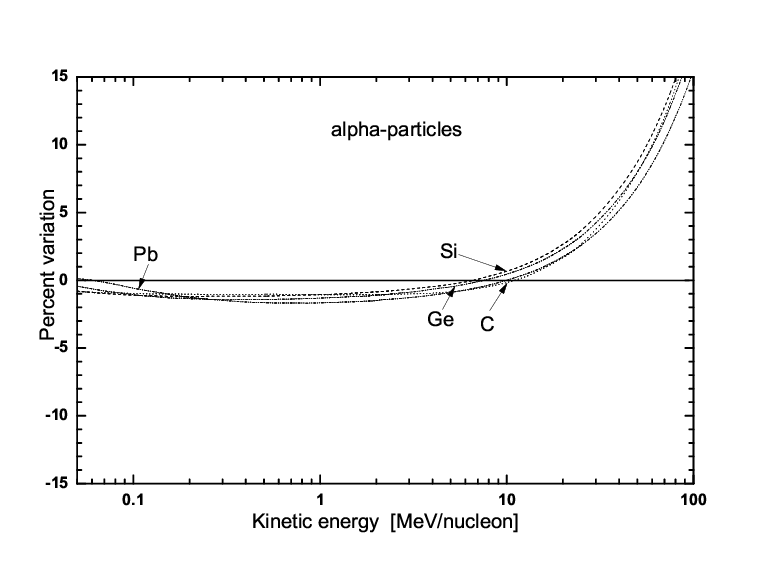,width=2.7in}
\end{center}
\vspace{-0.8cm}
\caption{Variation (in percentage) of nuclear stopping powers - calculated with Eq.~(\ref{de/dx_nuclear_T}) with the expressions~(\ref{eq:As_r_mod},~\ref{eq:C_mod}) for the screening parameter - with respect to ICRU tabulated values [3] as a function of the kinetic energy in MeV/nucleon, for $\alpha$-particles traversing amorphous carbon, silicon, germanium and lead media for energies from 50\,keV/nucleon up to 100\,MeV/nucleon.}
\label{fig:Var_As_mod}
\end{figure}
\par
For instance, in Fig.~\ref{fig:dEdx} (Fig.~\ref{fig:dEdx_Pb}) the nuclear stopping power in silicon (in lead) - in MeV\,cm$^2$\,g$^{-1}$ -
is shown as a function of the kinetic energy per nucleon - from 50\,keV/nucleon  up 100\,TeV/nucleon - for protons, $\alpha$-particles and $^{11}$B-, $^{12}$C-, $^{28}$Si-, $^{56}$Fe-, $^{115}$In-, $^{208}$Pb-nuclei.
\par
It has to be remarked that - at very low energies - the Wentzel--Moli\`{e}re nuclear stopping power [Eq.~(\ref{de/dx_nuclear_T})] differs from that obtained by Ziegler, Biersack and Littmark (1985) using the so-called \textit{universal screening potential} (see also Ref.~[18]).~However, they have shown (e.g.,~see Figure~2-18 in Ref.~[7] or, equivalently, Figure~2-18 in Ref.~[18]) that different screening potentials - including the Bohr potential in which $\Psi_{\rm I}(r_{\rm r})$ is assumed to be an exponential function similarly to Wentzel's assumption - result in nuclear stopping powers which exhibit marginal differences for $\epsilon_{\rm r,U}$ above 10 (see also Fig.~\ref{fig:ZBL}).~$\epsilon_{\rm r,U}$ is the so-called \textit{universal reduced energy} expressed as:
%
\begin{equation}\label{universal_red_energy}
   \epsilon_{\rm r,U}  =  \frac{\mathbb{R }}{z Z    \left( z^{0.23}+ Z^{0.23} \right) }  \left(\!\frac{M\,}{ m + M }\!\right)E_k ,
\end{equation}
where $E_k$ is in MeV and the numerical constant is $ \mathbb{R } = 32.536\! \times \!10^3$\,MeV$^{-1}$ \{e.g.,~see Equation~(2-73)
of~Ref.~[7] or Equation~(2-88) of~Ref.~[18], see also Section~2.1.4.1 of Ref.~[1]\}.~For instance, in silicon $\epsilon_{\rm r,U} \simeq 10$ corresponds to $E_k\simeq 13$\,keV [67\,keV/nucleon] for protons [lead nuclei].~Ziegler, Biersack and Littmark (1985) provided a general expression for the nuclear stopping power (e.g.,~see Section~2.1.4.1 of~Ref.~[1]), i.e.,
\begin{equation}\label{practical_NSP}
- \left(\frac{dE}{dx}\right)_{\rm nucl}^{\rm U} = \mathbb{K} \,\mathfrak{R}(\epsilon_{\rm
r,U})~~
\textrm{[MeV/cm]},
\end{equation}
with
\begin{equation}\label{practical_NSP_k_PARAMETER}
\mathbb{K} \simeq 5.1053\! \times \!
10^3 \, \frac{\rho\,zZ\,}{A\left(1+M/m\right) \left( z^{0.23}+ Z^{0.23} \right)}~~
\textrm{[MeV/cm]}
\end{equation}
where $\rho$ and $A$ are the density and atomic weight of the target medium,
respectively; $ \mathfrak{R} (\epsilon_{\rm r,U})$ - the
so-called \index{Nuclear!stopping power!reduced}\index{Reduced!nuclear stopping power|see{Nuclear, stopping power, reduced}}(\textit{universal}) \textit{reduced
nuclear stopping power} [termed, also, \index{Nuclear!stopping
power!scaled}(\textit{universal}) \textit{scaled nuclear stopping power}] given in Equations~(2-89)--(2-90) in Ref.~[18] (see also page~80 in~Ref.~[1]) - is dimensionless.~Additionally, the present calculations can be compared with values of stopping powers - obtained using the universal screened potential - available in SRIM~(2008) [19].~Usually an agreement - to better than a few percents - is achieved down to about 150\,keV/nucleon, where - for instance - one finds $\approx 5.5$ (9.9)\,\% for $\alpha$-particles (lead ions) in silicon.~At large energies, the non-relativistic approach due to Ziegler, Biersack and Littmark (1985) becomes less appropriate and deviations from stopping powers calculated by means of the universal screening potential are expected and observed for $\epsilon_{\rm r,U} \gtrsim (1.5$--$2.5) \times 10^{4}$ (e.g.,~see Fig.~\ref{fig:ZBL}).
\par
The non-relativistic approach - based on the universal screening potential - of Ziegler, Biersack and Littmark~(1985) was also used by ICRU~(1993) to calculate nuclear stopping powers - currently available on the web (e.g.,~see Ref.~[20]) - due to protons and $\alpha$-particles in materials.~ICRU~(1993) used as screening lengths those from Eqs.~(\ref{T_F_sc_rad},~\ref{Universal_sc_rad}) for protons and $\alpha$-particles, respectively.~In~Fig.~\ref{fig:Var}, the variation (in percentage) of nuclear stopping powers - calculated with Eq.~(\ref{de/dx_nuclear_T}) - with respect to ICRU tabulated values [3] is shown as a function of the kinetic energy per nucleon (in MeV/nucleon) - for energies from 50\,keV/nucleon up to 100\,MeV/nucleon - for protons and $\alpha$-particles traversing amorphous carbon, silicon, germanium and lead media.~The stopping powers for protons ($\alpha$-particles) from Eq.~(\ref{de/dx_nuclear_T}) are less
than $\approx 5$\% larger than those reported by ICRU~(1993) from 50\,keV/nucleon up to $\approx 32$\,MeV (31\,MeV/nucleon) - the upper energy corresponds to $\epsilon_{\rm r,U} \approx 6.2\times 10^{4}$ ($4.3\times 10^{4}$) for protons in an amorphous carbon ($\alpha$-particles in a silicon) medium -.~At larger energies the stopping powers from Eq.~(\ref{de/dx_nuclear_T}) largely differ from those from ICRU - as expected - due to the complete relativistic treatment of the present approach.
\par
The simple screening parameter used so far [Eq.~(\ref{eq:As_r})] - derived by Moli\`{e}re~(1947) - can be modified by means of a \textit{practical correction},~i.e.,
\begin{equation}\label{eq:As_r_mod}
A_{\rm s}'=\left(\frac{\hbar}{2\,p'_{in}\ a_{\rm I}}\right)^2\left[1.13+3.76 \times  \mathbb{C } \left(\frac{\alpha zZ}{\beta_{\rm r}}\right)^2\right],
\end{equation}
to achieve a better agreement with low energy calculations of Ziegler, Biersack and Littmark~(1985).~For instance, for protons, $\alpha$-particles and heavier ions, with
\begin{equation}\label{eq:C_mod}
\mathbb{C } = \left( 10 \pi z Z \alpha \right)^{0.04}
\end{equation}
the stopping powers obtained from Eq.~(\ref{de/dx_nuclear_T}) - in which $A_{\rm s}'$ replaces $A_{\rm s}$ - differ from the values of SRIM~(2008) by less than $\approx 3.5$ (2.6)\,\% for $\alpha$-particles (lead ions) in silicon down to about 50\,keV/nucleon.~With respect to the tabulated values of ICRU~(1993), the agreement for $\alpha$-particles is usually better than 2\% at low energy down to 50\,keV/nucleon (Fig.~\ref{fig:Var_As_mod}) - a 1\% agreement is achieved at about 50\,keV/nucleon in case of a carbon (or silicon) medium.~At very high energy, the stopping power is slightly affected when $A_{\rm s}'$ replaces $A_{\rm s}$: for example, a) in silicon at 100\,TeV/nucleon the nuclear stopping power of $\alpha$-particles (lead-ions) is decreased by about 0.03 (0.71)\,\% and b) in lead it is decreased by about 0.4 (1.0)\,\%.~It has to be remarked that a more appropriate expression for the screening parameter and a practical correction factor may require a further understanding.
\section{Non-Ionizing Energy Loss due to Coulomb Scattering}
A relevant process - which causes permanent damage to the silicon bulk structure - is the so-called \textit{displacement damage} (e.g.,~see Chapter~4 of Ref.~[1], Refs.~[21--23] and references therein).~Displacement damage may be inflicted when a \textit{primary knocked-on atom} (PKA) is generated.~The interstitial atom and relative vacancy are termed Frenkel-pair (FP).~In turn, the displaced atom may have sufficient energy to migrate inside the lattice and - by further collisions - can displace other atoms as in a collision cascade.~This displacement process modifies the bulk characteristics of the device and causes its degradation.~The total number of FPs can be estimated calculating the energy density deposited from displacement processes.~In turn, this energy density is related to the \textit{Non-Ionizing Energy Loss} (NIEL),~i.e., the energy per unit path lost by the incident particle due to displacement processes.
\begin{figure}[t]
\begin{center}
\vspace{-1.3cm}
\mbox{\hspace{-0.5cm}\psfig{file=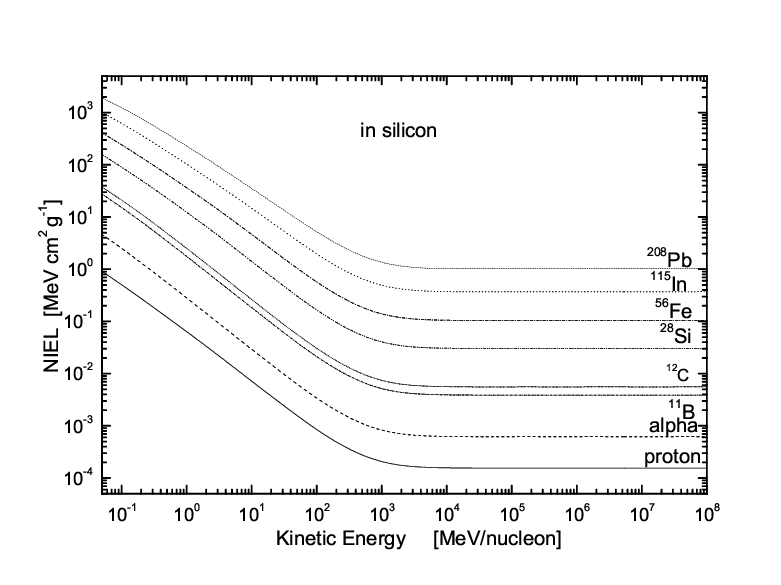,width=3.5in}}
\end{center}
\vspace{-0.8cm}
\caption{Non-ionizing stopping power - in MeV\,cm$^2$\,g$^{-1}$ - calculated using Eq.~(\ref{eq:NIEL}) in silicon is shown as a function of the kinetic energy per nucleon - from 50\,keV/nucleon  up 100\,TeV/nucleon - for protons, $\alpha$-particles and $^{11}$B-, $^{12}$C-, $^{28}$Si-, $^{56}$Fe-, $^{115}$In-, $^{208}$Pb-nuclei.~The threshold energy for displacement is 21\,eV in silicon.}
\label{fig:NIELdEdx}
\end{figure}
\begin{figure}
\begin{center}
\vspace{-1.3cm}
\mbox{\hspace{-0.5cm} \psfig{file=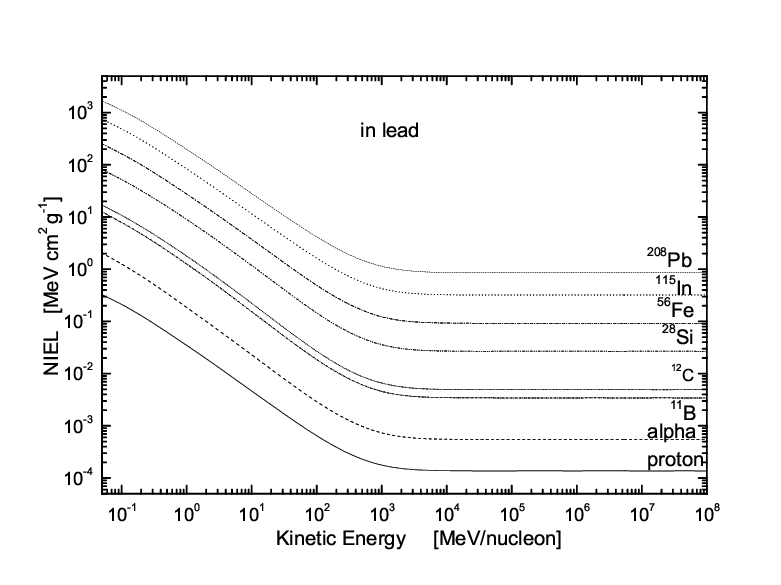,width=3.5in}}
\end{center}
\vspace{-0.8cm}
\caption{Non-ionizing stopping power - in MeV\,cm$^2$\,g$^{-1}$ - calculated using Eq.~(\ref{eq:NIEL}) in lead is shown as a function of the kinetic energy per nucleon - from 50\,keV/nucleon  up 100\,TeV/nucleon - for protons, $\alpha$-particles and $^{11}$B-, $^{12}$C-, $^{28}$Si-, $^{56}$Fe-, $^{115}$In-, $^{208}$Pb-nuclei.~The threshold energy for displacement is 25\,eV in lead}
\label{fig:NIELdEdx_Pb}
\end{figure}
\par
In case of Coulomb scattering on nuclei, the non-ionizing energy-loss can be calculated using the Wentzel--Moli\`{e}re differential cross section [Eq.~(\ref{eq:diff_cross_W_M_model_T_1})] discussed in Sect.~\ref{Nucleus_Nucleus_Potentials},~i.e.,
\begin{equation}\label{eq:NIEL}
- \left(\frac{dE}{dx}\right)_{\rm nucl}^{\rm NIEL}  = n_A\,\int^{T_{max}}_{T_d} \!T\,L(T)\,\frac{d\sigma^{\rm WM}(T)}{dT}\,dT\ ,
\end{equation}
where $E$ is the kinetic energy of the incoming particle, $T$ is the kinetic energy transferred to the target atom, $L(T)$ is the fraction of $T$ deposited by means of displacement processes.~The expression of $L(T)$ - the so-called \textit{Lindhard partition function} - can be found, for instance, in Refs.~[24,~25] and in Equations~(4.94,~4.96) of Section~4.2.1.1 in Ref.~[1] (see also references therein).~$T_{\rm de}= T\, L(T)$ is the so-called \textit{damage energy},~i.e.,~the energy deposited by a recoil nucleus with kinetic energy $T$ via displacement damages inside the medium.~The integral in Eq.~(\ref{eq:NIEL}) is computed from the minimum energy $T_d$  - the so-called \textit{threshold energy for displacement},~i.e., that energy necessary to displace the atom from its lattice position - up to the maximum energy $T_{max}$ that can be transferred during a single collision process.~$T_d$ is about 21\,eV in silicon (e.g., see Table~1 in Ref.~[26] and references therein) and 25\,eV in lead (e.g., see Table~22 at page~83 in Ref.~[27] and references therein).~For instance, in Fig.~\ref{fig:NIELdEdx} (Fig.~\ref{fig:NIELdEdx_Pb}) the non-ionizing energy loss - in MeV\,cm$^2$\,g$^{-1}$ - in silicon (lead) is shown as a function of the kinetic energy per nucleon - from 50\,keV/nucleon  up 100\,TeV/nucleon - for protons, $\alpha$-particles and $^{11}$B-, $^{12}$C-, $^{28}$Si-, $^{56}$Fe-, $^{115}$In-, $^{208}$Pb-nuclei.~As already remarked by Boschini and collaborators~(2010), at high energy the Coulomb NIEL - similarly to the nuclear stopping power - does not decrease with energy as it is found by Ziegler, Biersack and Littmark~(1985) or in other calculations based on their universal screening potential derived in the framework of a non-relativistic treatment of the screened Coulomb scattering.
\begin{figure}[t]
\begin{center}
\vspace{-0.6cm}
\psfig{file=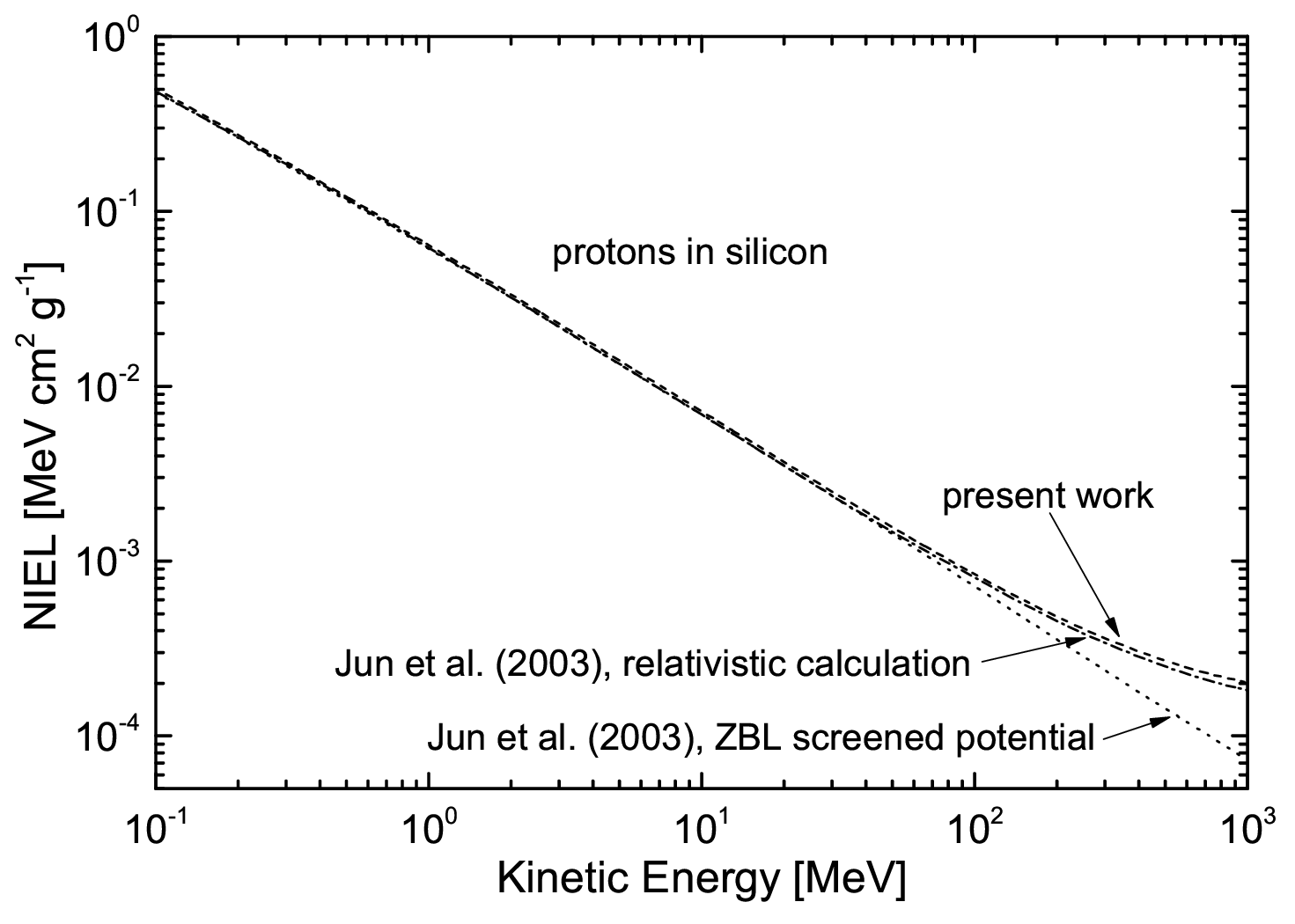,width=2.9in}
\end{center}
\vspace{-0.6cm}
\caption{Non-ionizing stopping powers of protons in silicon with energies from 100\,keV up to 1\,GeV: data for the dashed-dotted and dotted lines are the relativistic and Ziegler–-Biersack–-Littmark
screened potential calculations from Jun and collaborators~(2003), respectively (e.g.,~see Ref.~[26]); data for the dashed line are obtained from Eq.~(\ref{eq:NIEL}) (e.g, see Fig.~\ref{fig:NIELdEdx}). }
\label{fig:JUN}
\end{figure}
\par
Furthermore, Jun and collaborators~(2003) have already demonstrated that a relativistic treatment [24] of Coulomb scattering of protons - with kinetic energies above 50\,MeV - upon silicon results into a non-ionizing energy loss which is larger than that expected from calculations using the Ziegler--Biersack--Littmark screened potential with a universal screening length (e.g.,~see Refs.~[7,~25,~26]).~The relativistic cross section used for treating the Coulomb scattering is the one derived by McKinley and Feshbach~(1948) to describe the scattering of electrons on nuclei (e.g., see Section~4.2.1.4 of Ref.~[1] and references therein).~Seitz and Koehler~(1956) suggested that - when the mass of the projectile is much lower than the target rest-mass (e.g., see Section 13 of Ref.~[29] and references therein) - this cross section can also describe - although screening effects are neglected - the scattering of protons and light nuclei, thus, providing - at high energy - a damage cross section which does not decrease with increasing energy.~The data from Jun and collaborators~(2003) - for protons with energies from 100\,keV up to 1\,GeV - are shown in Fig.\ref{fig:JUN}: the Ziegler--Biersack--Littmark~(1985) screened potential was used to treat the Coulomb scattering of protons with energies lower than 50\,MeV.~In the same figure, the data obtained using Eq.~(\ref{eq:NIEL}) - e.g., see Fig.~\ref{fig:NIELdEdx} - are also shown.~There is an agreement to better than $\approx 6.5$\% - achieved at $\approx 1$\,GeV - between the results obtained by Jun and collaborators~(2003) and the present calculations.
\section{Conclusions}
The treatment of nucleus--nucleus interactions due to relativistic Coulomb scatterings with screened potentials - like the present approach based on Wentzel--Moli\`{e}re scattering - allows one to determine both the total and differential cross sections, thus, to calculate the resulting nuclear and non-ionizing stopping powers.~At high energies, the nuclear stopping powers exhibit a very slight logarithmic increase with energy.~At low energies - i.e., above $\approx 50$\,keV/nucleon and up to $\approx 32$ and 31\,MeV/nucleon, for protons and $\alpha$-particles, respectively -, the present results are in agreement to better than $\approx 5$\% with ICRU tabulated values of stopping powers obtained using the universal screening potential for the scattering of protons and $\alpha$-particles in matter.~An agreement to better than to (5.5--9.9)\,\% - for instance, for a silicon medium - is found with SRIM~(2008) values down to 150\,keV/nucleon.~Furthermore, these calculations are also in agreement to better than $\approx 6.5$\% with those obtained by Jun and collaborators~(2003) for the non-ionizing stopping powers of protons - with energies from 50\,MeV up to 1\,GeV - in silicon.
\par
Finally, it has to be remarked that - with a simple correction factor applied to the screening factor - an agreement to a few percents can be achieved with SRIM (2008) stopping powers down to about 50\,keV/nucleon.~However, a more appropriate expression for the screening parameter and a practical correction factor may require a further understanding.
\bibliographystyle{ws-procs9x6}
\bibliography{ws-pro-sample}

\end{document}